\documentclass[11pt]{article}

\usepackage{amssymb, amsmath, enumerate, theorem, epsfig}

\begin{document}

\mathchardef\bsurd="1371
\mathchardef\bsolid="132E

\def\dd#1#2{{{\rm d} #1\over {\rm d} #2}}
\def\ddn#1#2#3{{{\rm d}^#3 #1\over {\rm d} #2^#3}}
\def\bb#1{{\bf #1}}
\def\pd#1#2{\frac{{\partial #1}{\partial #2}}}
\def\bk{{\bf k}}
\def\bv{{\bf v}}
\def\bw{{\bf w}}
\def\bx{{\bf x}}
\def\bz{{\bf z}}

\def\pmb#1{\setbox0=\hbox{#1}%
\kern-.025em\copy0\kern-\wd0
\kern.05em\copy0\kern-\wd0
\kern-.025em\raise.0433em\box0}

\def\bze{{\pmb{$0$}}}
\def\bu{{\pmb{$1$}}}
\def\sm{{\pmb{$\cdot$}}}
\def\vm{{\pmb{$\times$}}}
\def\grad{{\pmb{$\nabla$}}}
\def\div{{\grad\sm}}
\def\curl{{\grad\vm}}
\def\bom{{\pmb{$\omega$}}}
\font\smallrm=cmr8 scaled \magstep 0

\title{\bf Pad\'e approximations of solitary wave solutions of the Gross-Pitaevskii
equation}
\author{Natalia G. Berloff\\
Department of Applied Mathematics and Theoretical Physics\\
University of Cambridge, Wilberforce Road, Cambridge, CB3 0WA, UK \\
{\it N.G.Berloff@damtp.cam.ac.uk}}
\date {Submitted 23 June 2003; in press by Journal of Physics A}
\maketitle
\begin {abstract} Pad\'e approximants are used to find 
approximate vortex solutions of any winding number in the context of Gross-Pitaevskii equation for a
uniform condensate and condensates with  axisymmetric
trapping potentials. Rational function and generalised rational function approximations of axisymmetric solitary waves of
the Gross-Pitaevskii equation are obtained in two and three dimensions. These approximations are
used to establish a new mechanism of vortex nucleation  as a result of
solitary wave interactions.
\end{abstract}

\section{Introduction}
In the last decade the experimental realisation of Bose-Einstein condensation in trapped 
alkali-metal gases at ultralow temperatures has stimulated an intense  
interest in the production of vortices and vortex arrays and theoretical 
investigations of their structure, energy, dynamics, and  stability 
\cite{fetter}. The
condensates of alkali vapours are pure and dilute, so that 
 the Gross-Pitaevskii (GP) 
model which represents the so-called mean-field
limit of quantum field theories gives a precise description of the
atomic condensates and their dynamics at low temperatures. The same equation has been the subject of
extensive studies also in the framework of superfluid helium at very low
temperature \cite{don}, though the high density and strong repulsive interactions
of superfluid helium restrict the applicability of the GP model so
that it provides at most a qualitative description.

In spite of the seeming simplicity of the GP model, which is a
defocussing 
nonlinear Schr\"odinger equation, that has also being extensively
studied in other physical systems such as nonlinear optics, not many asymptotic or
approximate solutions have been found especially for solitary wave
solutions in more than one dimensions. Typically one has to resort to  numerical integration
even in the case of a simple straight line vortex. Other techniques
involve  power series expansions that have a small radius of
convergence and, therefore, are  of a very limited use (see more
below). The other approach involves quite elaborate asymptotic
expansions in several different regions and asymptotic matching between
them (\cite{gr,br7}), which typically yields a useful result for  critical
parameters of interest such as  critical velocities, but the resulting approximation of the  solution is
too complicated to be used on its own, either in analytic manipulations or
as  an initial
condition (possibly with a perturbation) for numerical calculations.
Many numerical studies  use  a specific vortex
configuration as a starting initial condition (see for instance,
\cite{koplick,adams,nore,br9}) and it is desirable to have a simple
approximation to the vortex structure that can be used in such
calculations. 

The reason for the failure of the power series to represent the vortex
solution is that series diverge in the presence of
singularities. There are techniques of a summation theory that
allow us to overcome this difficulty and represent a given function by a
convergent expression. In Euler summation this expression is the limit
of a convergent series and in Borel summation this expression is the
limit of a convergent integral. But for these methods to work one has
to know all the terms of the divergent series exactly before the Euler
or Borel sum can be found approximately. Pad\'e summation  permits
us to use
only a few terms of the divergent series to construct an improved
estimate of the function.

In what follows we shall modify the standard Pad\'e summation method, so that only the general forms of the
power series at zero and at infinity are used to determine the
appropriate form of the  Pad\'e approximant, but the unknown
coefficients  will be determined recursively from the GP equation.
The general idea is that, if the function $f(x)$ has  power series
expansions of the form $x^n \sum_{i=1}^\infty p_i x^{2i-1}$ around
$x=0$ and $\sum_{i=0}^\infty q_i x^{-2i}$ at infinity, then a Pad\'e
approximation of the form
\begin{equation}
f(x)\approx \sqrt{\frac{x^{2n}\sum_{i=0}^N a_i
x^{2i}}{\sum_{j=0}^{N+n}b_jx^{2j}}},
\label{main}
\end{equation}
where $a_{N} = q_0^2b_{N+n}$ with  $q_0=f(\infty)$ will have the same
asymptotics at zero and infinity as the corresponding power series.

Our paper is organised as follows. In Section 2 we will derive a
Pad\'e approximation of the straight line vortex in a uniform condensate
for any winding number. Sections 3 and 4 deal with  vortices in a
condensate with trapping potentials including an external potential
with a laser beam. In Sections 5 and 6 we develop the rational and
generalised rational function approximations of the solitary waves
moving with a constant velocity in dimensions two and three
correspondingly. In Section 7 we study a new mechanism of  vortex
nucleation as a result of solitary wave interactions.
\section{Straight line vortex in a uniform condensate}

We start with  the GP model \cite{ginpit,gross} in the form
\begin{equation}
-2{\rm i} \frac{\partial\psi}{\partial t} = \nabla^2 \psi + (1 - |\psi|^2)\psi,
\label{GP}
\end{equation}
where we use dimensionless variables
 such that the unit of length corresponds to the healing
length $a$, the speed of sound is $c=1/\sqrt{2}$,  and the density at infinity is $\rho_\infty=1$.
The solution for the straight line vortex $\psi=R(r)\exp({{\rm i}n\theta})$ with winding number
$n=1,2,\cdot\cdot\cdot$ in a uniform condensate was first obtained by
Pitaevskii \cite{pit61} via 
numerical integration of the steady GP equation 
\begin{equation}
 R''(r) + \frac{1}{r}R'(r) -\frac{n^2}{r^2} R(r)+[1 - R^2(r)]R(r) = 0,
\label{steadyGP}
\end{equation}
subject to boundary conditions $R(0)=0$ and $R(\infty)=1$. The
asymptotic expansions for small $r$ and large $r$ in terms of the
power series were obtained by many authors typically for $n=1$.  At
small $r$, the solution can be asymptotically approximated by
$R(r)\sim\sum_{i=1}^{\infty}p_i r^{2i-1}$, where the first term of the
expansion $p_1$ has to be determined  numerically (by shooting) as
$p_1=0.5827811878$ with the rest of the terms then generated
recursively as $p_2=-0.072847648$, $p_3=0.01128249$,
$p_4=-0.001781398, \cdot\cdot\cdot$. The resulting series is useful
only within its radius of convergence (i.e., for $r<2.5$, \cite{naz}). Asymptotic but divergent solution at infinity
can be obtained as $R(r) \sim \sum_{i=0}^\infty q_i{r^{-2i}}$,
where $q_0=1$, $q_1=-\frac{1}{2}$, $q_2=-\frac{9}{8},
\cdot\cdot\cdot$. In view of the expansion in odd powers  at zero and
even power expansion at infinity a Pad\'e approximation of the straight line vortex
can be obtained in the form (\ref{main}) with $n=1$ and $N=1$ as
\begin{equation}
\rho(r) = R(r)^2={\frac{ r^2 (a_1
+ a_2  r^2)}{ 1 + b_1  r^2 + b_2  r^4}},
\label{exp}
\end{equation}
where we can let $b_2 = a_2$ in the view of the condition
$\rho\rightarrow 1$ as $r\rightarrow \infty$. We substitute the Pad\'e
approximation (\ref{exp}) into (\ref{steadyGP}) and expand the
resulting expression in series of $r$ setting the coefficient at equal
powers of $r$ to zero. At ${\cal O}(r^4)$ we get
$a_2 = a_1 (b_1 - 1/4)$, at ${\cal O}(r^6)$ we get $b_1=(5-32 a_1)/(48
- 192 a_1)$, and at  ${\cal O}(r^8)$ we get $a_1=11/32$ as the positive root of $11 +
56 a_1 - 256 a_1^2=0$. The resulting approximation 
\begin{equation}
R(r)\sim \sqrt{\frac{ r^2 (0.3437
+ 0.0286  r^2)}{ 1 + 0.3333 r^2 +  0.0286  r^4}}
\label{n1}
\end{equation}
gives the correct
asymptotic behaviour at $r\rightarrow 0$ and at $r\rightarrow \infty$
and 
 approximates the
numerical solution very well everywhere.
Figure 1 plots the various approximations to the numerically found
vortex solution for $n=1$. Note that the procedure described above
could be formalised by rescaling both  equation (\ref{steadyGP})
and the trial function (\ref{exp}) by $r \rightarrow \epsilon r$,
equating coefficients at powers of $\epsilon$ and setting $\epsilon=1$
(see also \cite{naz} for power series  expansions). This procedure will
become important when we consider solutions involving  more than one
variable, so that the recursive procedure can be established only by using
a different scaling of variables. We could also rewrite the equation
(\ref{steadyGP}) in terms of density $\rho = R^2(r)$, so that the
model equation for the straight line vortex becomes
\begin{equation}
\frac{d^2\rho}{dr^2} + \frac{1}{r}\frac{d\rho}{dr} -
\frac{1}{2\rho}\Bigr(\frac{d\rho}{dr}\Bigl)^2 - \frac{2\rho n^2}{r^2} - 2 \epsilon^2 (\rho -1)
\rho =0.
\label{rho}
\end{equation}

A Pad\'e
approximation of the straight line vortex with non unit winding number
can  be obtained by observing that $R(r)\sim r^{|n|}$ at $r\rightarrow 0$,
so that we need to consider the expression
\begin{equation}
\rho(r)\sim \frac{(\epsilon r)^{2|n|}(a_1
+ a_2 (\epsilon r)^2)}{\sum_{i=0}^{|n|}b_i(\epsilon r)^{2i} + a_2(\epsilon r)^{2|n|+2}}.
\label{nn}
\end{equation}
For instance, for $|n|=2$ we obtained
\begin{equation}
\rho\sim \frac{r^4(0.0256+ 
      0.0006264r^2)}{(1 + 0.19109 r^2 + 0.0196962 r^4 + 
    0.0006264 r^6)}.
\end{equation}
\begin{figure}
\centering
\caption{\baselineskip=10pt \footnotesize [Colour online] The plots of the amplitude
of the straight vortex line in a uniform condensate as a function of
distance from the centre of the vortex. The solid black line gives the
solution obtained by numerically integrating (\ref{steadyGP}). Grey
line -- the Pad\'e approximation (\ref{n1}). Short dashed and long
dashed lines give power series expansions at zero (to ${\cal
O}(r^{11})$) and at infinity (to ${\cal O}(r^{-20})$) correspondingly.}
\medskip
\psfig{figure=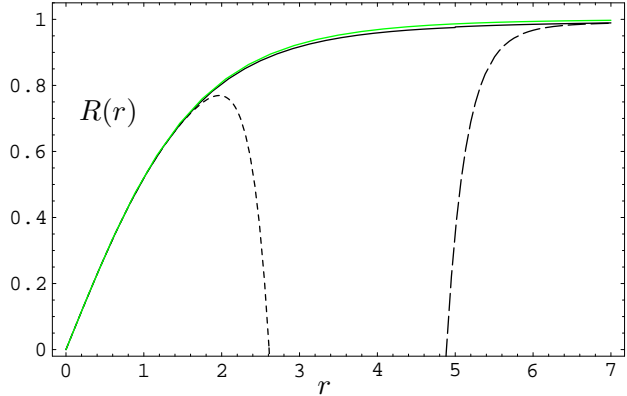,height=2.in}
\begin{picture}(0,0)(0,0)
\put(-120,-4) {$r$}
\put(-210,100){$R(r)$}
\end{picture}
\label{figure1}
\end{figure}

\section{Straight line vortex in a cigar-like
trap}
The equation for the density of a condensate in a cigar-like trap with
the vortex trapped in the centre is
\begin{equation}
\frac{d^2\rho}{dr^2} + \frac{1}{r}\frac{d\rho}{dr} - \frac{1}{2\rho}\Bigr(\frac{d\rho}{dr}\Bigl)^2 - \frac{2\rho}{r^2} - 2 \epsilon^2 (\rho -1+\epsilon^2\lambda^2 r^2)
\rho =0,
\label{rhocigar}
\end{equation}
where $\lambda$ is the dimensionless oscillator frequency.

For a noninteracting gas, the condensate wave function
for a singly quantised vortex on the symmetry axis involves the first
excited radial harmonic oscillator state $\psi_{\rm noninter}\sim r
\exp(-\frac{1}{ 2}\lambda^2 r^2) \exp(i \theta)$, so we seek
a solution of the form
\begin{equation}
\rho(r)={\frac{\epsilon^2 r^2 (a_1
+ a_2 \epsilon^2 r^2)}{1 + b_1 \epsilon^2 r^2 + b_2 \epsilon^4
r^4}}\exp(-\epsilon^2\lambda^2 r^2),
\label{cig}
\end{equation}
where the Taylor expansion will be taken for the exponential function. The
expression (\ref{cig}) with $b_2=a_2$ is substituted into
(\ref{rhocigar}) and 
 the terms up to ${\cal O}(\epsilon^{8})$ are set equal to zero. We
get 
\begin{eqnarray}
a_2&=&\frac{1}{4} a_1(-1 + 4\lambda^2 + 4 b_1), \\
b_1&=&-\frac{5+112
\lambda^4 - 48 \lambda^2 + 32 a_1(6\lambda^2 -1)}{ 48(4a_1 + 4 \lambda^2
-1)}
\end{eqnarray}
 with $a_1$ given as a positive root of
\begin{eqnarray}
256a_1^2(12\lambda^2-1)&+&(4\lambda^2-1)^3(52\lambda^2-11)\nonumber\\
&-&8a_1(384\lambda^6-528\lambda^4 + 136 \lambda^2-7)=0.
\end{eqnarray}
For example, the solution for $\lambda=0.2$ becomes
\begin{equation}\rho(r)=\frac{0.2833 r^2 + 0.0136 r^4}{ 1 + 0.2581 r^2 +
0.0136 r^4}\exp(-0.04 r^2).
\end{equation}

Similarly to the uniform condensate case, the approximation for the
multiply quantised vortices can be found in the form
\begin{equation}
\rho(r)\sim \frac{(\epsilon r)^{2|n|}(a_1
+ a_2 \epsilon^2 r^2)}{\sum_{i=0}^{|n|}b_i(\epsilon r)^{2i} +
a_2(\epsilon r)^{2|n|+2}}\exp[-\epsilon^2\lambda^2 r^2].
\label{nn2}
\end{equation}

\bigskip
\section{ Vortex in an axisymmetric condensate}
A similar procedure can be implemented to find the vortex in a
 condensate in an axisymmetric trap given by the external potential $V=\lambda^2 r^2 +
\lambda_Z^2 z^2$.
Now, the function $\rho$ depends on two coordinates $\rho=\rho(r,z)$,
so that
the equation for $\rho$ becomes
\begin{equation}
\rho_{rr} + \frac{\rho_r}{r} -
\frac{\rho_r^2}{2\rho} - \frac{2\rho}{r^2}+
 \frac{1}{ \epsilon^2}
\Biggl(\rho_{zz} - \frac{\rho_z^2}{2 \rho}\Biggr)- 2 \epsilon^2 (\rho
-1+\epsilon^2 \lambda^2 r^2 + \epsilon^4 \lambda_Z^2 z^2) \rho =0,
\label{axis}
\end{equation}
where we rescaled the variables as $r\rightarrow\epsilon r$ and
$z\rightarrow\epsilon^2 z$.
We seek a solution of the form
\begin{equation}
\rho(r,z)={\frac{\epsilon^2 r^2 (a_1
+ \epsilon^2 a_2  r^2)}{ 1 + \epsilon^2 b_1  r^2 + \epsilon^4 (b_2 
r^4 + c_1 z^2)}}\exp(-\epsilon^2\lambda^2 r^2-\epsilon^4\lambda_Z^2 z^2).
\label{rhorz}
\end{equation}

We solve the resulting equations to ${\cal O}(\epsilon^{8})$, define
all the parameters in terms of remaining two (say $c_1$ and $a_1$),
that are zeros of coupled polynomials, whose  roots can be
found numerically by fixing $\lambda$ and $\lambda_Z$. For instance,
for $\lambda=0.2$ and $\lambda_Z=0.1$ (and setting $\epsilon=1$) we have
\begin{equation}
\rho=\frac{0.232609 r^2 + 0.0319011 r^4}{1 + 0.342901 r^2 +
0.0319011  r^4 + 0.00697428 z^2}\exp(-0.04 r^2-0.01 z^2).
\label{solaxis}
\end{equation}
Figure 2 gives the contour plot of this solution in $xz-$plane.
The generalisation to multiply charged vortices is carried out in a
similar way to 
previous sections.

The asymptotics of vortices in  a trapped condensate was studied by Konotop
and Perez-Garcia \cite{konotop} using a sophisticated multiscale
method; their solution was of
the form $A \tanh(a r) \exp(-b r^2 - c z^2)$, where $A,a,b,c$ are some
constants that needed to be estimated numerically. The use of a Pad\'e
approximation not only provides a swift and easy tool for generating an
approximation of comparable accuracy but also has  the ability to include a
$z-$dependence that is not confined to the exponential term.
\begin{figure}
\centering
\caption{\baselineskip=10pt \footnotesize A density contour plot in the
$xz-$plane for a condensate containing a vortex along the
$z-$axis as the solution of (\ref{axis}) given by (\ref{solaxis}). The trap parameters are $\lambda=0.2$ and
$\lambda_Z=0.1$. Luminosity is proportional to density, the white area
being the most dense.}
\medskip
\psfig{figure=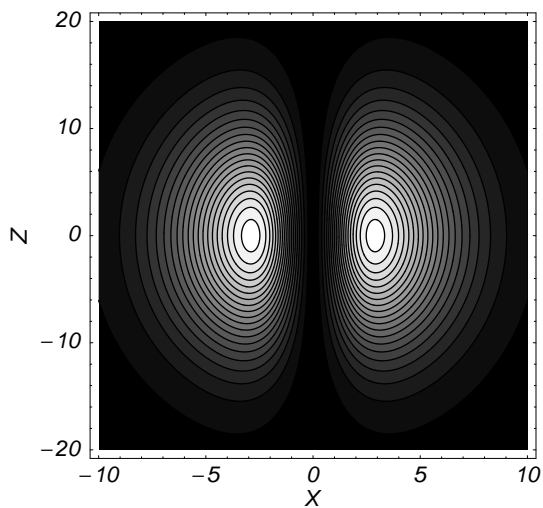,height=2.8in}
\label{figure2}
\end{figure}

There are different ways of creating a vortex in a trapped
condensate. In particular,  angular momentum  can be transmitted to the
condensate by rotationally stirring it 
with a laser beam
\cite{caradoc} or by guiding a vortex created at the edge of the
condensate by a laser beam towards the centre of condensate
\cite{staliunas}. When the vortex is brought to 
rest at the centre of the  condensate, the steady external potential can be
assumed to have  the form 
\begin{equation}
V(r,z)=\lambda^2 r^2 + \lambda_Z z^2 + V_0 \exp[-r^2/r_l^2],
\label{laser}
\end{equation}
where $r_l$ is the half-width of the laser beam intensity profile.
Our procedure for finding the approximate solution can be
automatically adjusted for finding the vortex density in the
condensate with potential (\ref{laser}). We seek a
Pad\'e approximant (\ref{rhorz}) as a solution of  the equation
\begin{eqnarray}
\rho_{rr} + \frac{\rho_r}{r} -
\frac{\rho_r^2}{2\rho} &-& \frac{2\rho}{r^2}+
 \frac{1}{ \epsilon^2}
\Biggl(\rho_{zz} - \frac{\rho_z^2}{2 \rho}\Biggr)\nonumber \\
&-& 2 \epsilon^2 (\rho
-1+\epsilon^2 \lambda^2 r^2 + \epsilon^4 \lambda_Z^2 z^2+ V_0 \exp[-\epsilon^2r^2/r_l^2]) \rho =0.
\label{axis2}
\end{eqnarray}
In particular, for $V_0=r_l=0.8$ we get the following solution
\begin{equation}
\rho=\frac{0.18354 r^2 + 0.0910069 r^4}{1 + 0.452004 r^2 +
0.0910069  r^4 + 0.005353 z^2}\exp(-0.04 r^2-0.01 z^2).
\label{solaxis2}
\end{equation}
Figure 3 shows the $r-$dependent plots of (\ref{solaxis}) and (\ref{solaxis2}) for
various values of $z$. The form of the external potential is given as an
inset. Notice how the laser beam causes a slight depletion of the
condensate close to the centre, followed by an increase in the density elsewhere.
\begin{figure}
\centering
\caption{\baselineskip=10pt \footnotesize Plots of the density
function $\rho(r,z)$ at $z=0,5,10,15$
for the vortex solutions with ((\ref{solaxis2}) dashed line) and
without ((\ref{solaxis}) solid line) the laser beam. The form of the
external potential is depicted in the inset.}

\medskip
\psfig{figure=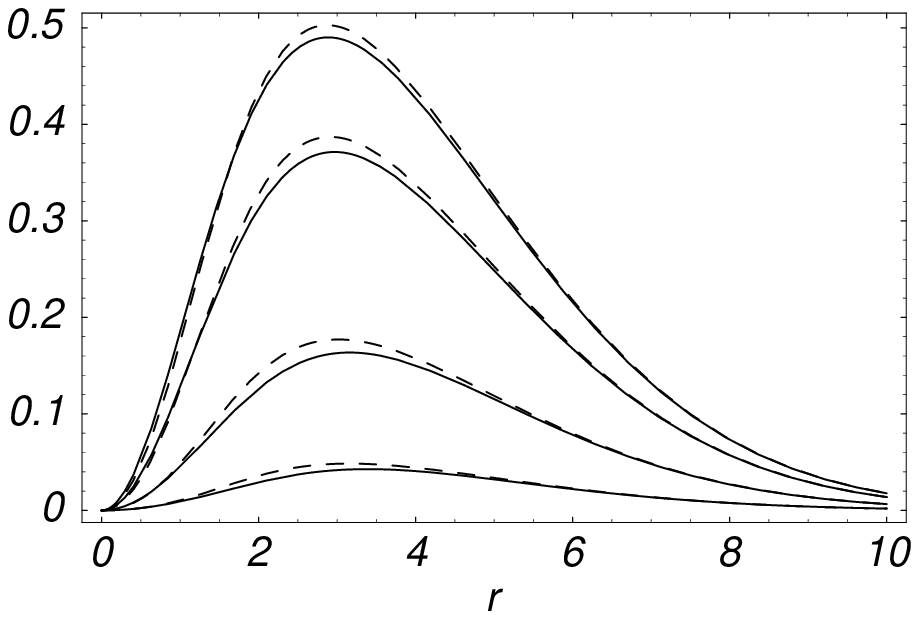,height=3in}
\begin{picture}(0,0)(0,0)
\put(-170,120){{\psfig{figure=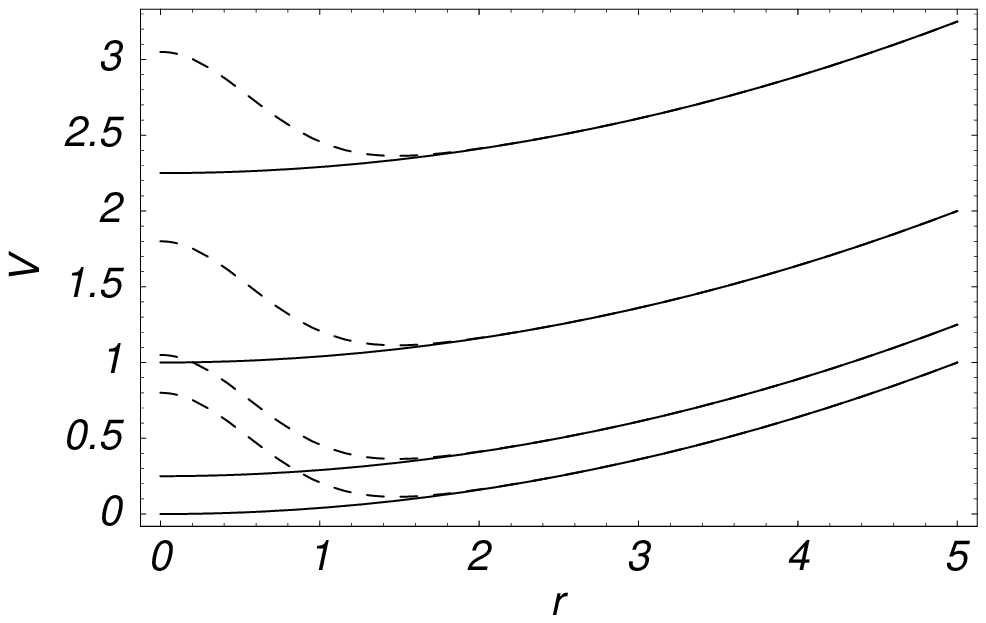,height=1.2in}}}
\put(-280,200){$\rho(r,0)$}
\put(-230,180){$\rho(r,5)$}
\put(-210,105){$\rho(r,10)$}
\put(-210,60){$\rho(r,15)$}
\put(-100,195){\footnotesize $z=15$}
\put(-100,170){\footnotesize $z=10$}
\put(-120,150){\footnotesize $z=5$}
\put(-80,138){\footnotesize $z=0$}
\end{picture}
\label{figure3}
\end{figure}
\section{Pad\'e approximations of solitary waves in two dimensions}
So far we have shown that  Pad\'e approximations are useful for
obtaining  accurate approximations of  straight line vortices. In
this section we obtain  approximate solutions of  solitary waves 
 moving with a constant velocity.
In a seminal paper, Jones
and Roberts \cite{jr} numerically  determined the entire sequence
of solitary wave solutions of the GP equation, such as vortex rings,
vortex pairs, and rarefaction pulses. Their numerics involved the
introduction of stretched variables, expansion of the wave function in double
Chebyshev-Legendre series and a Newton-Raphson iteration of the
resulting system of nonlinear algebraic equations.

We start with finding Pad\'e approximations of solitary waves in two
dimensions (2D).
The pair of two point vortices of opposite circulation centred at
$(0,y_0)$ and $(0,-y_0)$ with large $y_0$ can be
approximated by the superposition of wavefunctions of two
straight-line vortices of opposite circulation, therefore, by the field with the density
\begin{equation}
\rho =
R^2(\sqrt{x^2+(y-y_0)^2})R^2(\sqrt{x^2+(y+y_0)^2}),
\label{rhoplot}
\end{equation}
 and the phase
\begin{equation}
S=\arctan\Biggr[\frac{y-y_0}{x}\Biggl]-\arctan\Biggr[\frac{y+y_0}{x}\Biggl]
\label{ssplot},
\end{equation}
where $R(r)$ is given by (\ref{n1}).
Figure 4 illustrates how  this approximates the solution obtained
numerically by a Newton-Raphson iteration of the GP equation

\begin{equation}
2 i U \frac{\partial \psi}{\partial z} = \nabla^2 \psi + (1 -
|\psi|^2)\psi,
\label{U}
\end{equation}
subject to the boundary condition $\psi \rightarrow 1$ as $|\bb
x|\rightarrow \infty$. Here $U$
is the
velocity with which the vortices of opposite circulation propel each
other in the positive $x-$direction.
\begin{figure}
\centering
\caption{\baselineskip=10pt \footnotesize Plots of the density function $\rho(0,y)$
for vortex pairs with separations $2y_0=8$ (4a) and $2y_0=1.78$ (4b) taken
in a cross-section through the centres of vortices. Solid line -- the
approximate solution obtained by multiplying the wave functions of
individual vortices, dashed line -- solution obtained numerically.}

\medskip
\hbox{\psfig{figure=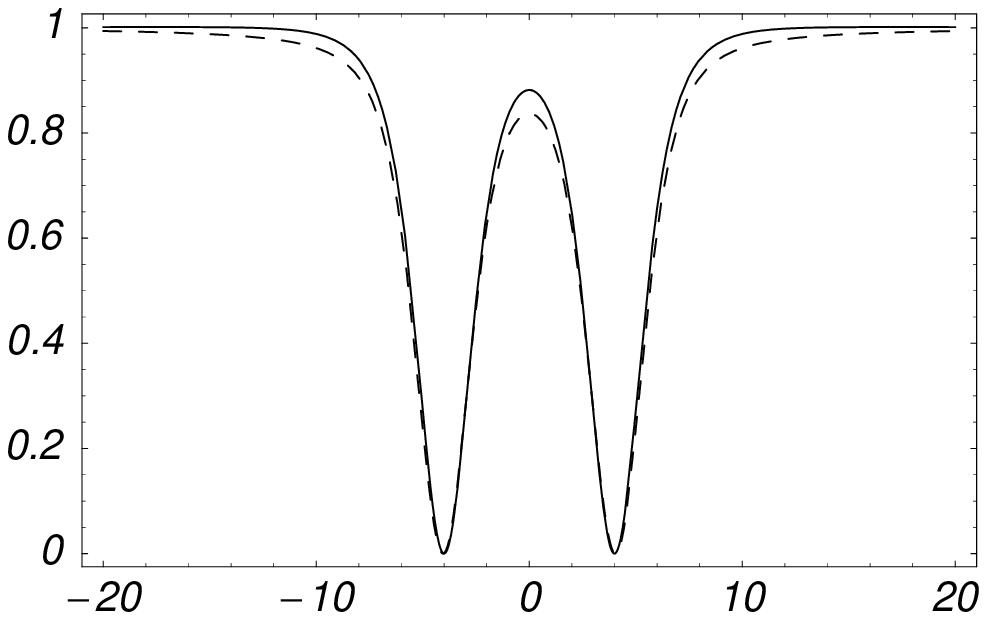,height=1.8in}\psfig{figure=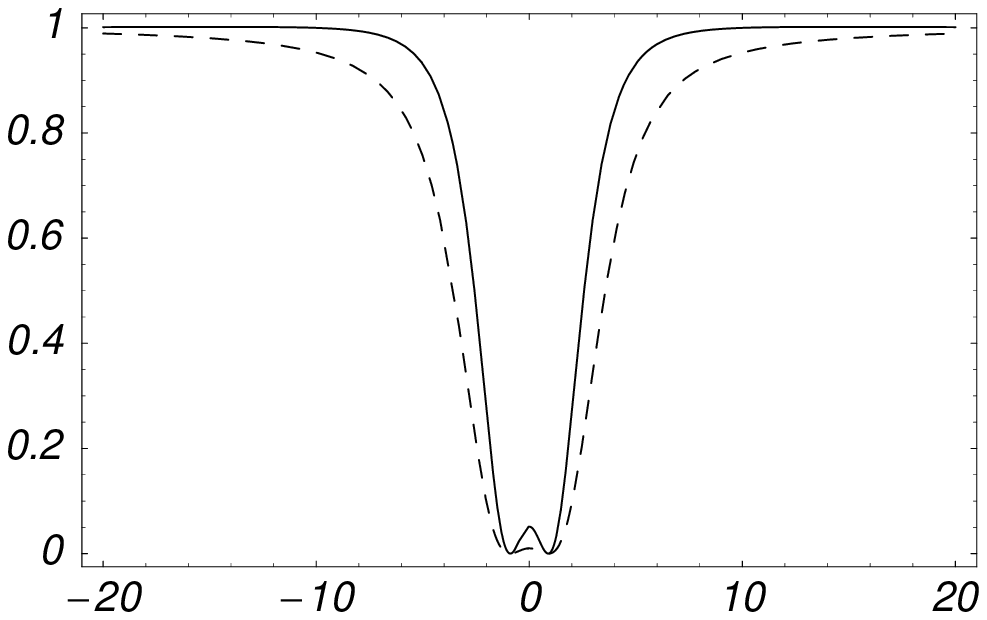,height=1.8in}}
\begin{picture}(0,0)(0,0)
\put(-40,120) {4(a)}
\put(170,120){4(b)}
\put(-100,5) {$y$}
\put(-180,125){$\rho(0,y)$}
\put(110,5) {$y$}
\put(30,125){$\rho(0,y)$}
\end{picture}
\label{figure4}
\end{figure}

The approximation (\ref{rhoplot})-(\ref{ssplot}) is sufficiently
accurate for large $y_0$, but significantly deviates from the
numerical solution as the separation between the vortices decreases. Another
difficulty in using this approximation is that one needs to know the separation $2y_0$ for each $U$ in order
to construct such an approximation.

To obtain Pad\'e approximations of 2D solitary solutions we observe
that the approximation (\ref{rhoplot})-(\ref{ssplot}) written for the
real $u(x,y)$ and imaginary $v(x,y)$ parts of the wavefunction
$\psi(x,y)=\sqrt{\rho}\exp[-{\rm i} S] =u(x,y) + {\rm i} v(x,y)$ are
\begin{eqnarray}
u(x,y)&=&(x^2+y^2-y_0^2)\tilde R(\sqrt{x^2+(y-y_0)^2})\tilde R(\sqrt{x^2+(y+y_0)^2}),\nonumber\\
v(x,y)&=&-2 x y_0 \tilde R(\sqrt{x^2+(y-y_0)^2})\tilde R(\sqrt{x^2+(y+y_0)^2}),
\label{uv0}
\end{eqnarray}
where $\tilde R(r)=R(r)/r.$
This suggests that a Pad\'e approximation of the solitary wave solution
moving in $x-$direction with a constant velocity $U$ can be
found in the form
\begin{eqnarray}
u(x,y)&=&\frac{\sum a_{ij} x^{2 i}y^{2j}}{1+\sum
  c_{ij} x^{2 i}y^{2j}},\nonumber\\
v(x,y)&=&x\frac{\sum b_{ij} x^{2 i}y^{2j}}{1+\sum
  c_{ij} x^{2 i}y^{2j}}.
\label{uv_int}
\end{eqnarray}
We truncate (\ref{uv_int}) to the lowest order that can give two zeros of
the density and unity at infinity, so we seek an
approximation of the form 
\begin{eqnarray}
u(x,y)&=&1+\frac{ a_{00}+a_{10}x^2 + a_{01}y^2}{1+c_{10}x^2 +
  c_{01}y^2 + c_{20}x^4 +c_{11}x^2y^2+ c_{02}y^4},\nonumber\\
v(x,y)&=&x\frac{b_{00}+b_{10}x^2 + b_{01}y^2}{1+c_{10}x^2 +
  c_{01}y^2 + c_{20}x^4 +c_{11}x^2y^2+ c_{02}y^4}.
\label{uv}
\end{eqnarray}
We fix the value of $U$, substitute (\ref{uv}) into the equation
(\ref{U}), expand in powers of $x$ and $y$ and set the coefficients at
the first three leading orders to zero. As the result we get 12
algebraic equations in 11 variables $a_{ij},b_{ij},$ and $c_{ij}$ that
are compatible and
can be solved by a computer algebra. The   first two leading orders
give the analytical expressions of six coefficients in terms of
remaining five. The sum of squares of the remaining six equations is
further numerically minimised on the set of remaining five coefficients.

For $U=0.4$ the solution was found as
\begin{eqnarray}
u&=&1+\frac{-1.10048 - 0.095006\,x^2 + 0.01681\,y^2}
  {1+\,0.3108\,x^2 + 0.0192\,x^4 + 0.1030\,y^2 +
  0.0219\,x^2\,y^2 + 6.112\times 10^{-3}\,y^4}
,\nonumber\\
v&=&\frac{x\,\left( -0.818647 - 0.06910\,x^2 - 0.03756\,y^2 \right) }
  {1 + 0.3108\,x^2 + 0.0192\,x^4 + 0.1030\,y^2 +  0.0219\,x^2\,y^2 +6.112\times 10^{-3}\,y^4}.
\label{u0.4}
\end{eqnarray}
\begin{figure}
\centering
\caption{\baselineskip=10pt \footnotesize The equidistant contour and
  density plots
  of
  the density $\rho=u(x,y)^2+v(x,y)^2$ where $u(x,y)$ and $v(x,y)$ are
  given by (\ref{u0.4b}). Luminosity is proportional to density, the white area
being the most dense.} 
\medskip
\hbox{\psfig{figure=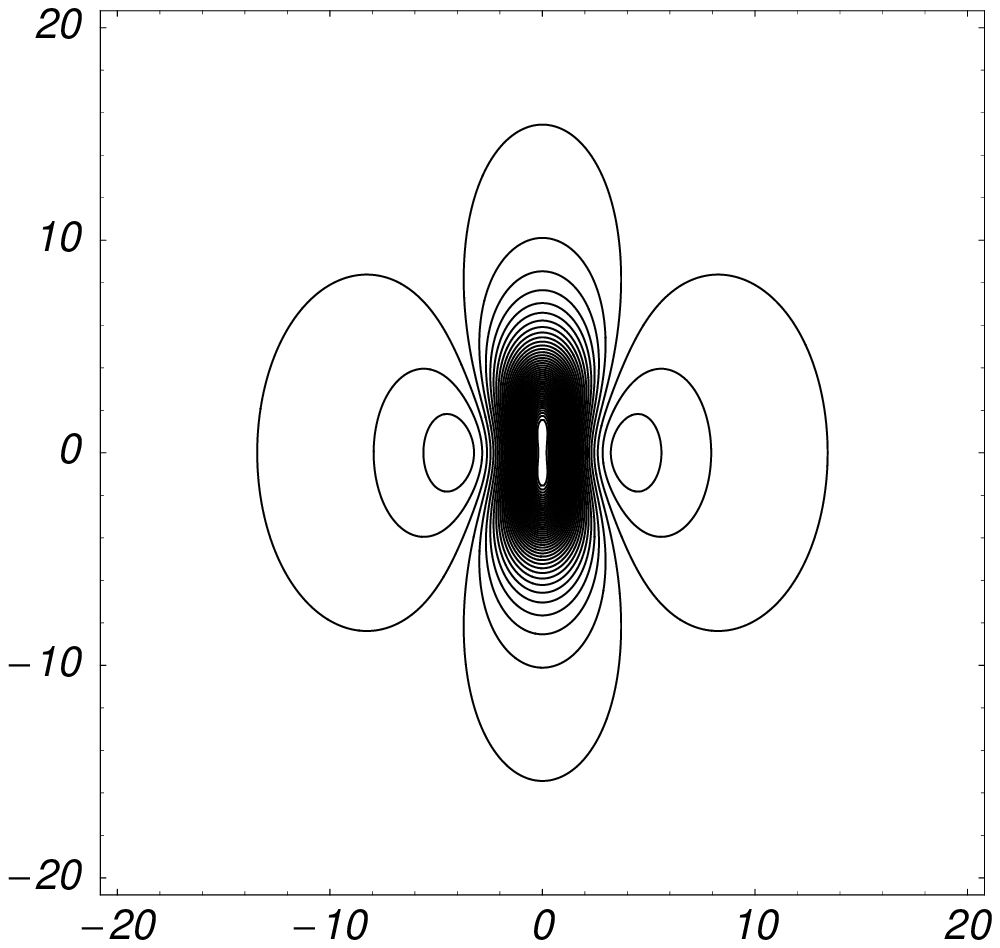,height=1.8in}\psfig{figure=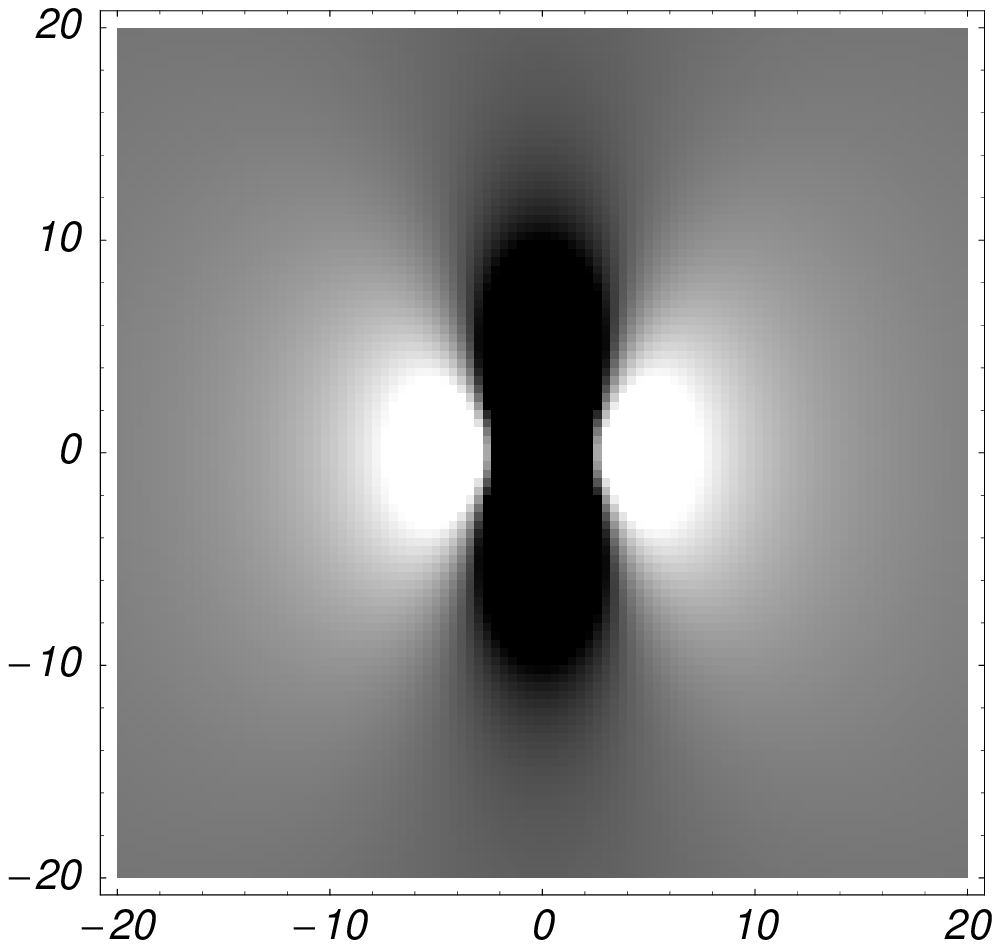,height=1.8in}}
\label{fig4}
\end{figure}
The error  of the approximation (\ref{u0.4}) is $Err_1 \approx 0.008$,
where we defined
\begin{equation}
{Err_1} =\max_{x,y}(2 Uv_x +\nabla^2u+ (1
  - u^2-v^2)u)^2 +(-2 U u_x +\nabla^2v + (1
  - u^2-v^2)v)^2.
\label{err1}
\end{equation}
From our constructions the expressions (\ref{u0.4}) approximate the solution very well
for small $x$ and $y$ and give  unity for the density at infinity.
Two zeros of $u(0,y)$ give the half separation between vortices as
$0.893997 $ (compare it with  the value found numerically in \cite{jr} as
$0.89$ to two significant digits).

We can improve the error $Err_1$, by considering not just the global
minimum of the sum of the squares of ${\cal O}(\epsilon^4)$ error in
the power series expansions about the origin, but
  also  local minima and choosing the one that minimises
  $Err_1$. This procedure gives us an approximate solution with
  $Err_1<10^{-4}$ which is
\begin{eqnarray}
u&=&1+\frac{-1.09077 - 0.0983212\,x^2 - 0.00193044\,y^2}
  {1 + 0.31406\,x^2 + 0.02026\,x^4 + 0.11677\,y^2 + 0.02448\,x^2\,y^2 + 0.007432\,y^4}
  ,\nonumber\\
v&=&\frac{x\,\left( -0.811702 - 0.0717083\,x^2 - 0.0465306\,y^2 \right) }
  {1 + 0.314063\,x^2 + 0.02026\,x^4 + 0.11677\,y^2 + 0.024476\,x^2\,y^2 + 0.007432\,y^4}.
\label{u0.4b}
\end{eqnarray}
The energy, ${\cal E,}$ and momentum, $p,$ of solitary wave solutions  can be calculated as in \cite{jr5}: 
\begin{eqnarray}
{\cal E}&=&\tfrac{1}{4}\int(1 - u^2-v^2)(3-2u-u^2-v^2)dx\,dy,\nonumber\\
Up&=&\tfrac{1}{2}\int (1 - u^2-v^2)^2dx\, dy,
\label{ep}
\end{eqnarray}
and are ${\cal E}\approx 8.1$ and $p\approx 14.2$ for $u$ and $v$
given by (\ref{u0.4b}). These values can be compared with ${\cal E}
\approx 8.16$ and $p\approx 14.1$ found numerically in
\cite{jr}. Since we relaxed the precision of the approximation for
small $x$ and $y$, the  half separation between vortices lost its
precision as well, becoming $0.87$.
Figure 5 shows the contour and density  plots of the density
$\rho(x,y)=u(x,y)^2+v(x,y)^2$ of the approximate solution (\ref{u0.4b}). 

Finally we consider the third approximation, where we further relax
the precision of the approximation  at the origin, but  impose  additional
constraints at large $x$ and $y$. The asymptotic solution for large
$x$ and $y$ was obtained in \cite {jr} as
\begin{eqnarray}
u(x,y)&\approx&1 + \frac{m(U-\tfrac{1}{2} m)x^2 -mU(1-2U^2)y^2}{(x^2 + (1-2
  U^2)y^2)^2},\nonumber\\
v(x,y)&\approx&-\frac{mx}{x^2 + (1-2
  U^2)y^2},\quad\quad |\bf x|\rightarrow \infty,
\label{asy}
\end{eqnarray}
where $m$ is a constant termed `the stretched dipole moment' of the
wave because of the factor $1-2U^2$.
If we would like our solution to have  this asymptotic behaviour at infinity, we have to let 
\begin{eqnarray}
a_{10}&=&c_{20}m(U-\tfrac{1}{2} m), \quad a_{01}=-c_{20} m(1 - 2
U^2)U, \quad b_{10}=-m c_{20}\nonumber\\
 b_{01}&=&-m c_{20} (1-2 U^2), \quad c_{11}=2  c_{20} (1-2U^2), \quad c_{02}=c_{20} (1-2U^2)^2.
\label{cond}
\end{eqnarray}
The  expansion of
the equation (\ref{U}) about zero to ${\cal O}(\epsilon^2)$ gives four  equations on 
seven unknowns $U, m, a_{00}, b_{00}, c_{20}, c_{10}, c_{01}$. We can
solve these equations analytically for  $c_{20}, c_{10}, c_{01},$ and
$b_{00}.$ In particular, the following approximate relation between the minimum
of the real part of the wavefunction $u(0,0)=a_{00}+1$ and the slope of
the imaginary part at the origin in the direction of a solitary motion
$v_x(0,0)=b_{00}$ was
found
\begin{equation}
v_x(0,0)=(u(0,0)-1)\frac{16 + m\,U - 20\,U^4}{7\,\left( m - 4\,U^3 \right)
}.
\label{ba}
\end{equation}
Next
we fix one parameter, say $U$, to specify one particular solution of
the family,
 and find $a_{00}$ and $m$ that minimise the integral error
\begin{equation}
Err_2=\int(2 Uv_x +\nabla^2 u+ (1
  - u^2-v^2)u)^2 +(-2 U u_x +\nabla^2 v + (1
  - u^2-v^2)v)^2\, dV,
\label{err2}
\end{equation}
where $dV=dx \,dy$.

By implementing this procedure, we obtained the following
approximation of the vortex pair moving with  $U=0.4$
\begin{eqnarray}
u(x,y)&=& 1+\frac{-1.14026 - 0.150112\,x^2 - 0.0294564\,y^2}
  {1 + 0.35022\,x^2 + 0.03032\,x^4 + 0.15905\,y^2 + 0.04123\,x^2\,y^2 + 0.01402\,y^4},
\nonumber \\
v(x,y)&=&\frac{x\,\left( -0.830953 - 0.108296\,x^2 - 0.073641\,y^2 \right) }
  {1 + 0.35022\,x^2 + 0.03032\,x^4 + 0.15905\,y^2 +
  0.04123\,x^2\,y^2 + 0.01402\,y^4}
\label{u0.4c}
\end{eqnarray}
with the stretched dipole moment found as $m=3.57$ (compared with
the value
$3.55$ found numerically in \cite{jr}). The absolute error of this
approximation is $Err_1 \approx 0.0018$.

In order to find the solitary wave with a single zero of the wave
function, we implement the latter procedure by fixing $a_{00}=-1$. This
will guarantee that the intersection of zeros of real and imaginary
parts is only at the origin. We found that $U=0.45$, $m=3.32$ and
the approximate solution with $Err_1=0.00089$ is
\begin{eqnarray}
u(x,y)&=& 1+\frac{-1 - 0.137233\,x^2 - 0.0303092\,y^2}
  {1 + 0.36838\,x^2 + 0.03397\,x^4 + 0.15803\,y^2 + 0.04082\,x^2\,y^2 + 0.01227\,y^4}
\nonumber \\
v(x,y)&=&\frac{x\,\left( -0.803361 - 0.112913\,x^2 - 0.0678506\,y^2 \right) }
  {1 + 0.36838\,x^2 + 0.03397\,x^4 + 0.15803\,y^2 + 0.04082\,x^2\,y^2 + 0.01227\,y^4}.
\label{u0}
\end{eqnarray} 
On Figure 6 the contour  plot of $\rho(x,y)$
is given. This solution is interesting because it represents the
borderline case between vortex pair and vortex-free solutions.

\begin{figure}
\centering
\caption{\baselineskip=10pt \footnotesize Equidistant contour plot
of the density $\rho=u(x,y)^2+v(x,y)^2$ where $u(x,y)$
and $v(x,y)$are given by (\ref{u0}).} 
\medskip
Figure was not included in this submission to keep the size below the maximum allowed.
\label{figure5}
\end{figure}

A Pad\'e approximation of a rarefaction pulse can be obtained
similarly. For instance, if $U=0.5$ we get $a_{00}=-0.826$ and
$m=3.1$, so that the solution is
\begin{eqnarray}
u(x,y)&=& 1+\frac{-0.825937 - 0.114393\,x^2 - 0.0271467\,y^2}
  {1 + 0.37355\,x^2 + 0.03495\,x^4 + 0.14235\,y^2 + 0.03495\,x^2\,y^2 + 0.008737\,y^4},
\nonumber \\
v(x,y)&=&\frac{x\,\left( -0.737901 - 0.108587\,x^2 - 0.0542934\,y^2 \right) }
  {1 + 0.37355\,x^2 + 0.03495\,x^4 + 0.14235\,y^2 + 0.03495\,x^2\,y^2 + 0.008737\,y^4}.
\label{rare}
\end{eqnarray} 
We will use this approximation of the rarefaction pulse in Section 7, where we  study  vortex nucleation.
\section{Generalised rational function approximation of axisymmetric
  solitary waves in three dimensions}
To determine approximations of the axisymmetric vortex rings and
rarefaction pulses 
moving along the $x-$axis with the constant velocity $U$ in three
dimensions (3D),
we need to solve (\ref{U}) with the Laplacian written in cylindrical
coordinates
\begin{eqnarray}
-2Uv_x&=&u_{xx} + u_{ss} + \tfrac{1}{s}u_s + (1 - u^2-v^2)u,\nonumber \\
2Uu_x&=&v_{xx} + v_{ss} + \tfrac{1}{s}v_s + (1 - u^2-v^2)v,
\label{UU}
\end{eqnarray}
where $s=\sqrt{y^2+z^2}$.
The asymptotics of the solitary waves at large distances from the
origin was found in \cite{jr} to have the form 
\begin{eqnarray}
u(x,s)&\approx&1 + \frac{2mUx^2 - 2mU(1-2U^2)s^2}{(x^2 +
  (1-2U^2)s^2)^{5/2}},\nonumber \\
v(x,s)&\approx&-\frac{mx}{(x^2 +
  (1-2U^2)s^2)^{3/2}},\quad\quad |\bf
x|\rightarrow \infty.
\label{assring}
\end{eqnarray}
It is clear that approximation by rational functions can not behave at
large distances as (\ref{assring}), therefore, we will use `generalised'
rational functions. In particular,
we determined that among many possibilities the following expressions
 give a reasonable approximation everywhere
\begin{eqnarray}
u(x,y)&=&1+\frac{ a_{00}+a_{10}x^2 + a_{01}s^2+m c_{20}^{7/4}U(2x^2-(1
  - 2U^2)s^2)}{(1+c_{10}x^2 +
  c_{01}s^2 + c_{20}(x^2+(1-2U^2)s^2)^2)^{7/4}},\nonumber\\
v(x,y)&=&x\frac{b_{00}+b_{10}x^2 + b_{01}y^2-mc_{20}^{7/4}(x^2 + (1 -
  2U^2) s^2)^2}{(1+c_{10}x^2 +
  c_{01}s^2 + c_{20}(x^2+(1-2U^2)s^2)^2)^{7/4}}.
\label{pade}
\end{eqnarray}
From the two leading order expansions of (\ref{UU}) about the origin  we determine $a_{10},a_{01},b_{01},b_{10}$ and the rest of the unknowns will be found by
minimising (\ref{err2}), where $dV=s\, ds\, dx$, 
 for a fixed $U$. Among many local minima that give small $Err_2$,  we choose the one with a
 stretched dipole moment close to the one obtained in \cite{jr}. In
 particular, the small vortex ring with $U=0.6$ and $Err_1=0.003$ was
 found as (\ref{pade}) with
\begin{eqnarray}
a_{00}&=&-1.1,\quad a_{01}=0.0170524, \quad a_{10}=0.0289452, \quad
m=8.97 \nonumber \\
b_{00}&=&-0.953, \quad b_{01}=-0.0049767, \quad b_{10}=-0.0594346, \\
c_{01}&=&0.04, \quad c_{10}=0.21, \quad c_{20}=0.00612.\nonumber
\label{vr}
\end{eqnarray} 
We calculate the energy and momentum of the solitary wave solutions in
3D as (see \cite{jr})
\begin{eqnarray}
{\cal E}&=&\pi\int(u_x^2+u_s^2+v_x^2+v_s^2+\tfrac{1}{2}(1 - u^2-v^2)^2)s\,ds\,dx,\nonumber\\
p&=&{2\pi}\int ((u-1)v_x-vu_x)s\, ds\, dx,
\label{ep3d}
\end{eqnarray}
to get ${\cal E}\approx 58.8$ and $p\approx78$  for (\ref{pade})
with (\ref{vr}), that can be compared with the
values found numerically in \cite{jr}: ${\cal E}\approx 56.4$ and
$p\approx 78.9$. The radius of the ring given by (\ref{pade})-(\ref{vr}) is $1.059$
(with 1.06 found numerically). 

Finally we give  our approximation of  the
rarefaction pulse moving with $U=0.63$ found on the lower branch of the dispersion curve
calculated numerically in \cite{jr}:
\begin{eqnarray}
a_{00}&=&-0.79792,\quad a_{01}=0.00388059 , \quad a_{10}=0.00882276, \quad
m=8.37 \nonumber \\
b_{00}&=&-0.7981, \quad b_{01}=-0.012783, \quad b_{10}=-0.0574092 ,\nonumber \\
c_{01}&=&0.0399, \quad c_{10}=0.199, \quad c_{20}=0.0058.
\label{rare3d}
\end{eqnarray} 
For this approximation of the rarefaction pulse (\ref{pade})-(\ref{rare3d}) we have
  ${\cal
  E}\approx 54.4$, $p\approx72.2$, and $m=8.37$ (compared with numerical
  values found in \cite {jr}: ${\cal
  E}\approx 52.3$, $p\approx72.2$, and $m=8.37$).
\section{Vortex nucleation}
\begin{figure}[h!t]
\centering
\caption{\baselineskip=10pt \footnotesize [Colour online] The snapshots of the 
contour plots of the density cross-section of a condensate obtained by numerically
integrating the GP model (\ref{GP}). Initial condition is
$\psi(t=0)=\Psi(x-5,0)\Psi(x+5,0)$, 
where $\Psi=u+{\rm i} v$, with $u$ and $v$ given by
(\ref{pade})-(\ref{rare3d}). Black solid lines show zeros of real and
imaginary parts of $\psi$, therefore, their intersection shows the
position of topological zeros.  Both low and high density regions are
shown in 
darker shades to emphasise  intermediate density regions. Only a portion of an actual computational box is shown.}
\medskip
\vskip 0.2 in
\psfig{figure=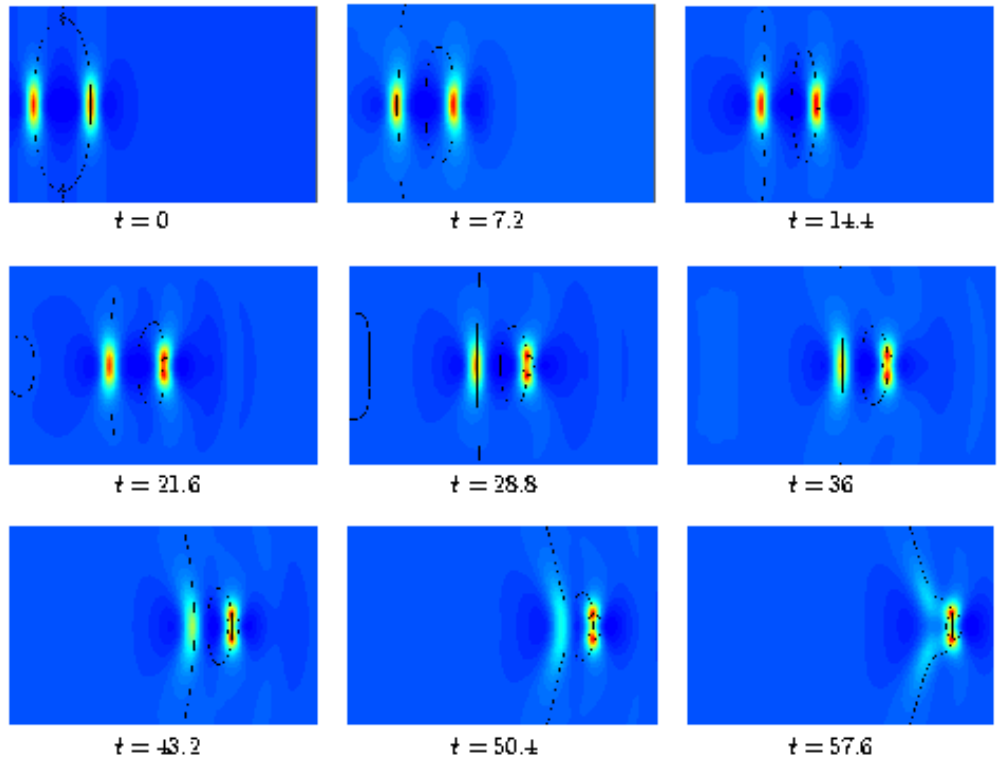,height=4.in}
\label{fig6}
\end{figure}
\begin{figure}[h!]
\centering
\caption{\baselineskip=10pt \footnotesize [Colour online] The isosurface
  $\rho/\rho_{\infty}=0.15$ of a condensate. 
 Initial condition is
$\psi(t=0)=\Psi(x-5,0)\Psi(x+5,0)$, 
where $\Psi=u+{\rm i} v$, with $u$ and $v$ given by
(\ref{pade})-(\ref{rare3d}). At $t=0$ the solution is vortex free. AT $t=22.5$ the
  front pulse contains a closed topological zero. At $t=40$ a
  well-formed vortex ring has developed with the back solution
  becoming too shallow to be shown for this isosurface. 
Only a portion of an actual computational box is shown.}
\medskip
\vskip 0.2 in
\psfig{figure=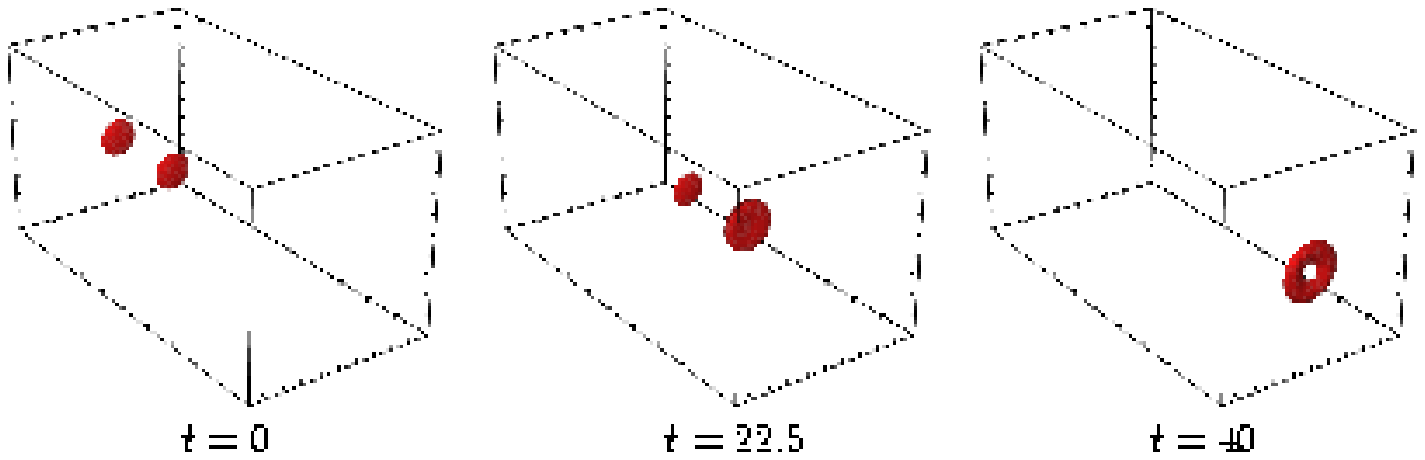,height=2.in}
\label{fig7}
\end{figure}

The approximations developed in the previous sections allow one to
study interactions among the solitary wave
solutions of the GP equation. Without an accurate starting point in
numerical calculations it would be impossible to separate clearly the effect
of interactions from the evolution of each  solitary wave  by itself as it
settled down from a poor initial guess.

In this section we will use the  approximations developed above
to show that  vortex pairs and vortex rings can appear as a result
of an interaction among the solitary wave solutions of the GP
equation. Previously, the nucleation of vortices in a uniform
condensate  has been connected to
 critical velocities \cite{frisch,br7} or instabilities of the
initial states \cite{berloff}. We will show that  interactions
between various, even vortex-free, solitary waves result in  energy and
momentum transfer  that can lead to 
vortex nucleation. Rarefaction pulses on the lower branch of the dispersion curve
have lower energy and momentum than vortex rings, therefore, such
rarefaction pulse may evolve into a vortex ring if interactions with
other solutions add enough energy and momentum to the rarefaction pulse.

This scenario is supported by direct numerical simulations, performed
with the same numerical method as in our previous work
\cite{br7,br9}. In these computations we follow the evolution of two
rarefaction pulses  moving in the
computational box of dimensions $D^3=80^3$. The faces of the box are
open to allow sound waves to escape; this is achieved numerically by
applying the Raymond-Kuo technique \cite{rk}.

We prepare our initial conditions by superimposing the wavefunctions $\psi_i$
of solitary wave solutions of the GP equation found in the previous
sections,
$\psi(t=0)=\prod\psi_i$. If the distance between such solutions is large
enough, then such a superposition will not  lead to a significant
initial 
sound emission and solitary waves will preserve their form initially.
Our first initial state consists of
two rarefaction pulses (\ref{pade})-(\ref{rare3d}) positioned the
distance 10 apart  and moving towards each other. Two pulses collide and pass
through each other 
without loss of energy. Next we will create a field nonuniformity
by placing two rarefaction pulses a distance 10 apart that move in the same direction.
This time the effect  two solitary waves have on each other is
non-symmetric. As a result, the rarefaction pulse moving behind
transfers part of its energy and momentum to the pulse moving at
front, so that the latter transforms into a vortex ring and slows down,
whereas the former  spreads out and speeds
up. This process leads to an  even closer interaction of the two solitary
waves and an even more rapid transfer of energy from the solitary wave that
moves behind to the one moving at front. Eventually almost all of the
energy and momentum of the former is transferred to the latter, which
 becomes a vortex ring of energy and momentum that are only slightly
less than twice the
energy and momentum of each of the initial rarefaction pulses. The
remaining small energy is emitted as sound waves. Figure 7 gives the
graphical illustration of this process through the snapshots of the
density cross-sections. Figure 8 shows the density isoplots at
$|\psi|^2=0.15$ of the various stages of the ring formation.

Similarly,  energy and momentum transfer takes place between
different types of solitary waves. In our next calculation we start
with a rarefaction pulse followed by a large vortex
ring that moves in the same direction. Initially both the 
distance between these two solitary waves  and the radius of the
ring were taken to be 10. 
The rarefaction pulse is moving  faster
than the vortex ring, so the distance between them  rapidly
increases. Nevertheless, there is an energy and momentum transfer
that allows the rarefaction pulse to evolve into a vortex ring of a
small radius and the large vortex ring to shrink slightly. Apart from
a small loss of energy to sound waves, the total energy of these two
solitary waves is almost conserved throughout this transformation. These processes are shown in graphical form through the 
contour plots of the density cross-sections (Figure 9) and the
density isoplots at $|\psi|^2=0.1.$ (Figure 10).
\begin{figure}[h!]
\centering
\caption{\baselineskip=10pt \footnotesize [Colour online] The snapshots of the 
contour plots of the density cross-section of a condensate obtained by numerically
integrating the GP model (\ref{GP}). The initial condition is
$\psi(t=0)=\Psi_1(x,0)\Psi_2(x-5,s+10)\Psi_2^*(x-5,s-10)$, 
where $\Psi_1=u+{\rm i} v$, with $u$ and $v$ given by (\ref{pade})-
(\ref{rare3d}) and $\Psi_2=R(\sqrt{x^2+y^2})e^{{\rm i} \theta}$ with $R$
given by (\ref{n1}). Black solid lines show zeros of real and
imaginary parts of $\psi$, therefore, their intersection shows the
position of topological zeros.  Both low and high density regions are
shown in 
darker shades to emphasise intermediate density regions. Only a portion of an actual computational box is shown.}
\medskip
\vskip 0.2 in
\psfig{figure=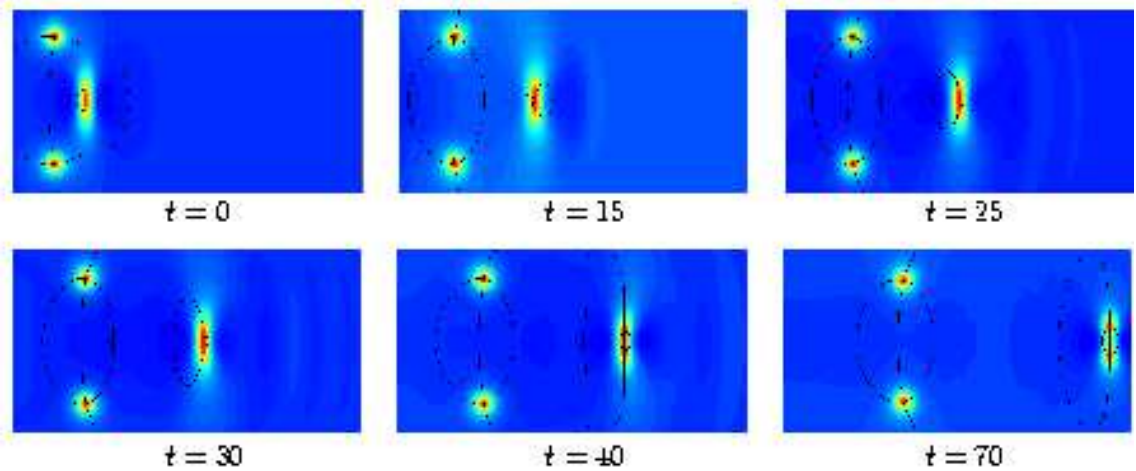,height=2.5in}
\label{fig8}
\end{figure}
\begin{figure}[h!]
\centering
\caption{\baselineskip=10pt \footnotesize [Colour online] The isosurface
  $\rho/\rho_{\infty}=0.1$ of a condensate. 
 The initial condition is
$\psi(t=0)=\Psi_1(x,0)\Psi_2(x-5,s+10)\Psi_2^*(x-5,s-10)$, 
where $\Psi_1=u+{\rm i} v$, with $u$ and $v$ given by (\ref{pade})-
(\ref{rare3d}) and $\Psi_2=R(\sqrt{x^2+s^2})e^{{\rm i} \theta}$ with $R$
given by (\ref{n1}).
 At $t=0$ the front pulse is vortex free. AT $t=22.5$ the
  front pulse contains a closed topological zero. At $t=40$ the front pulse evolved into a
  well-formed vortex ring.
Only a portion of an actual computational box is shown.}
\medskip
\vskip 0.2 in
\psfig{figure=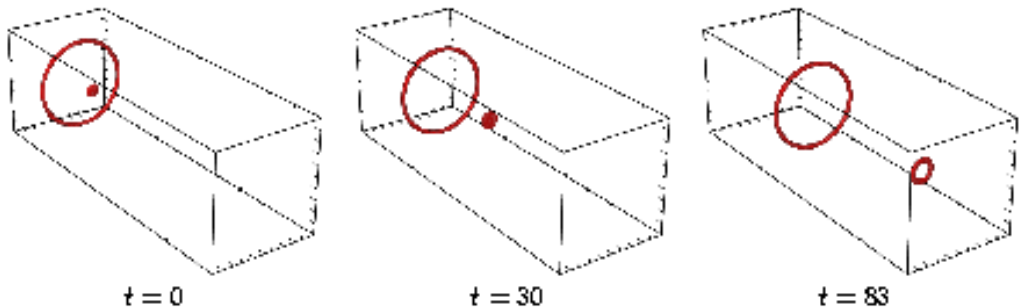,height=1.7in}
\label{fig9}
\end{figure}

Similar transfer of energy and momentum from one solitary wave to
 another takes place in 2D. In particular, there is a transfer of energy between
 solitary waves moving in the same direction, whereas  two colliding
 solitary waves interact elastically. Nevertheless, this elasticity of interactions
 can be broken by introducing other solitary waves. As an example, we
 show the evolution of an initial condition consisting of two
 colliding rarefaction pulses in close vicinity of a widely separated vortex pair. 
 As a result of  the interaction, another vortex pair is created
and the resulting two pairs of vortices move apart in direction making
 small angles
with the  positive $x-$ axis. Figure 11 shows the snapshots of the density
of a condensate at various moments of time as the solution evolves.
\begin{figure}[h!]
\centering
\caption{\baselineskip=10pt \footnotesize [Colour online] The snapshots of the 
contour plots of the density of a condensate obtained by numerically
integrating the GP model (\ref{GP}) in 2D. Initial condition is
$\psi(t=0)=\psi_1(x-10,0)\psi_1^*(x+10,0)\psi_2(x,y+4)\psi_2^*(x,y-4)$, 
where $\psi_1=u+{\rm i} v$, with $u$ and $v$ given by
(\ref{rare}) and $\psi_2=R(\sqrt{x^2+y^2})e^{{\rm i} \theta}$ with $R$
given by (\ref{n1}). Black solid lines show zeros of real and
imaginary parts of $\psi$, therefore their intersection shows the
position of topological zeros. Both low and high density regions are
shown in 
darker shades to emphasise  intermediate density regions.  Only a portion of an actual computational box is shown.}
\medskip
\vskip 0.2 in
Figure was not included in electronic submission to keep the size below maximum allowed.
\label{fig10}
\end{figure}

\section{Conclusions}
In summary we have presented a new technique for finding  approximate
vortex solutions of the GP equation in a uniform condensate and in
condensates with axisymmetric traps. These solutions have simple
analytic expressions, correct asymptotic behaviours at zero and infinity
and approximate the entire solutions quite well elsewhere. We envision
that the use of such approximations will allow one to set up accurate
initial vortex configurations for numerical calculations and will make
explicit analytic 
manipulations possible.

We have also developed a technique for obtaining  approximations of the solitary solutions such as
vortex pairs and rarefaction pulses in two and three dimensions. The found approximations are
shown to give a very low error and have  simple analytical
form. These approximations are used to elucidate 
the energy and momentum transfer between different solitary wave solutions
by direct numerical integration of the GP equation. The process of
vortex nucleation is one of the consequences of such a transfer.

\section*{Acknowledgements}
The author is grateful to Professor Paul Roberts for useful
discussions and comments about this manuscript.
This work is supported by the NSF grant DMS-0104288.

\end{document}